\documentclass[11pt]{elsarticle}
\usepackage{geometry}
\usepackage[english]{babel}
\usepackage[utf8]{inputenc}
\usepackage{graphicx}
\usepackage{glossaries}
\usepackage{amsmath}
\usepackage{xfrac}
\usepackage{verbatim}
\usepackage{listings}
\usepackage{amssymb}
\usepackage{latexsym}
\usepackage{amsthm}
\usepackage{eucal}
\usepackage{float}
\usepackage{longtable}
\usepackage{subcaption}

\title{Search for dark matter annihilation in the Sun using the full ANTARES dataset}

\date{}

\usepackage{hyperref}
\counterwithin{figure}{section}
\hypersetup{
 colorlinks=true,        
 linkcolor=cyan,          
 citecolor=cyan,        
 filecolor=cyan,      
 urlcolor=cyan          
}

\begin{document}


\begin{frontmatter}
\author[IPHC,UHA]{A.~Albert}
\author[IFIC]{S.~Alves}
\author[UPC]{M.~Andr\'e}
\author[UPV]{M.~Ardid}
\author[UPV]{S.~Ardid}
\author[CPPM]{J.-J.~Aubert}
\author[APC]{J.~Aublin}
\author[APC]{B.~Baret}
\author[LAM]{S.~Basa}
\author[APC]{Y.~Becherini}
% ...

\address[IPHC]{\scriptsize{Universit\'e de Strasbourg, CNRS,  IPHC UMR 7178, F-67000 Strasbourg, France}}
\address[UHA]{\scriptsize Universit\'e de Haute Alsace, F-68100 Mulhouse, France}
\address[IFIC]{\scriptsize{IFIC - Instituto de F\'isica Corpuscular (CSIC - Universitat de Val\`encia) c/ Catedr\'atico Jos\'e Beltr\'an, 2 E-46980 Paterna, Valencia, Spain}}
\address[UPC]{\scriptsize{Technical University of Catalonia, Laboratory of Applied Bioacoustics, Rambla Exposici\'o, 08800 Vilanova i la Geltr\'u, Barcelona, Spain}}
\address[UPV]{\scriptsize{Institut d'Investigaci\'o per a la Gesti\'o Integrada de les Zones Costaneres (IGIC) - Universitat Polit\`ecnica de Val\`encia. C/  Paranimf 1, 46730 Gandia, Spain}}
\address[CPPM]{\scriptsize{Aix Marseille Univ, CNRS/IN2P3, CPPM, Marseille, France}}
\address[APC]{\scriptsize{Universit\'e Paris Cit\'e, CNRS, Astroparticule et Cosmologie, F-75013 Paris, France}}
\address[LAM]{\scriptsize{Aix Marseille Univ, CNRS, CNES, LAM, Marseille, France }}
% ...
\end{frontmatter}


\section*{Abstract}
Weakly Interacting Massive Particles (WIMPs) are among the most popular candidates for particle dark matter.
In certain scenarios, WIMPs could be gravitationally captured into massive celestial objects and then annihilate into Standard Model particles, possibly leading to the emission of  neutrinos. 
The ANTARES neutrino telescope, located in the Mediterranean Sea, is sensitive to neutrinos with energies ranging from a few tens of GeV to the TeV scale, making it well suited to search for the annihilation of WIMPs with masses below a few TeV/c$^2$.
One of the closest potential astrophysical sources of neutrinos from WIMP annihilations is the Sun. 
In this work, a search for high-energy neutrinos originating from the Sun, using the full dataset recorded by ANTARES from 2007 to 2022 was conducted. The analysis yielded no significant evidence for a neutrino excess over the expected background, and upper limits on the WIMP-proton interaction cross section have been obtained, for both spin-dependent and spin-independent processes. These limits are valid for dark matter masses ranging from 35\,GeV/c$^2$ to 10\,TeV/c$^2$ for the annihilation benchmark channels WIMP$+$WIMP $\longrightarrow$ $b\bar{b},\tau^+\tau^-, W^+W^-$, assuming a 100$\%$ branching ratio for each of these channels. These results improve on those previously obtained by the ANTARES Collaboration by a factor of two-three and are consistent with those reported by other experiments in direct and indirect searches for dark matter.


\section{Introduction}

The presence of \textit{non-baryonic} matter in the Universe has been hypothesised to explain different astrophysical and cosmological observations. From the study of the dynamics of galaxy clusters \cite{zwicky} and of the rotational curves of spiral galaxies \cite{rubin}, to the analysis of the power spectrum of temperature anisotropies in the Cosmic Microwave Background (CMB) \cite{cmb}, and the measurement of matter density in large celestial bodies such as galaxies and galaxy clusters via gravitational lensing \cite{lensing}, many observations point towards the existence of a non-luminous, collisionless component in the matter content of the Universe, called dark matter (DM). In particular, under the $\Lambda$CDM cosmological model, the analysis of  CMB data shows that the amount of dark matter is five times larger than that of ordinary matter \cite{CMB1, CMB2}.\\
Several extensions of the Standard Model (SM) of particle physics have been proposed to explain the dark matter content of the Universe. 
Any DM particle candidate must be electromagnetically neutral, and should only interact with ordinary matter via the gravitational force and possibly via interactions that are at most of the same order of the SM weak force ($\le 10^{-42}$\,cm$^2$/nucleon) \cite{susy}. 
In order to detect and study these candidates, different experimental approaches are possible \cite{Bertone}: producing DM particles at colliders \cite{collider}; directly detecting DM interactions with nuclei in underground laboratories \cite{direct}; indirectly detecting DM by observing cosmic messengers (cosmic rays, gamma rays, neutrinos) originated by its interactions or decays in the Universe \cite{indirect}. 
Indeed, Weakly Interacting Massive Particles (WIMPs) may annihilate into short-lived SM particles whose decays can produce secondary particles, such as neutrinos, that can be detected on Earth.
The associated signatures have been extensively searched for at neutrino telescopes \cite{rewDM}. In particular, the ANTARES Collaboration has searched for DM signals from the Galactic Centre \cite{ANTARESGC, ANTGCSDM}, the Sun \cite{SunC, ANTSunSDM}, and the Earth \cite{ANTARESEarth}. The Sun represents a particularly interesting target because of the small astrophysical background; indeed, the flux of high-energy (GeV - TeV) neutrinos produced by cosmic-ray interactions in or near the Sun has already been constrained by several instruments, including ANTARES itself \cite{daniel}, yielding no significant detection of this background flux. \\
In this work, a search for neutrinos originating from WIMP annihilation in the core of the Sun is presented, using data recorded by the ANTARES neutrino telescope from 2007 to 2022, increasing by a factor of three the amount of data analysed with respect to previous searches \cite{SunC, ANTSunSDM}.
The paper is organised as follows: the production of neutrinos from DM annihilation in the Sun is discussed in Section~\ref{sec:sun}. The ANTARES detector and the different event reconstruction strategies used are described in Section~\ref{sec:antares}. In Section~\ref{sec:analysis} the analysis strategy is described and the results are shown in Section~\ref{sec:results}. Finally, the conclusions are presented in Section~\ref{sec:conclusions}.

\section{The Sun as a source of neutrinos from dark matter annihilation}
\label{sec:sun}

\begin{figure}[!h]
\centering

\begin{subfigure}{0.46\textwidth}
    \centering
    \includegraphics[width=\linewidth]{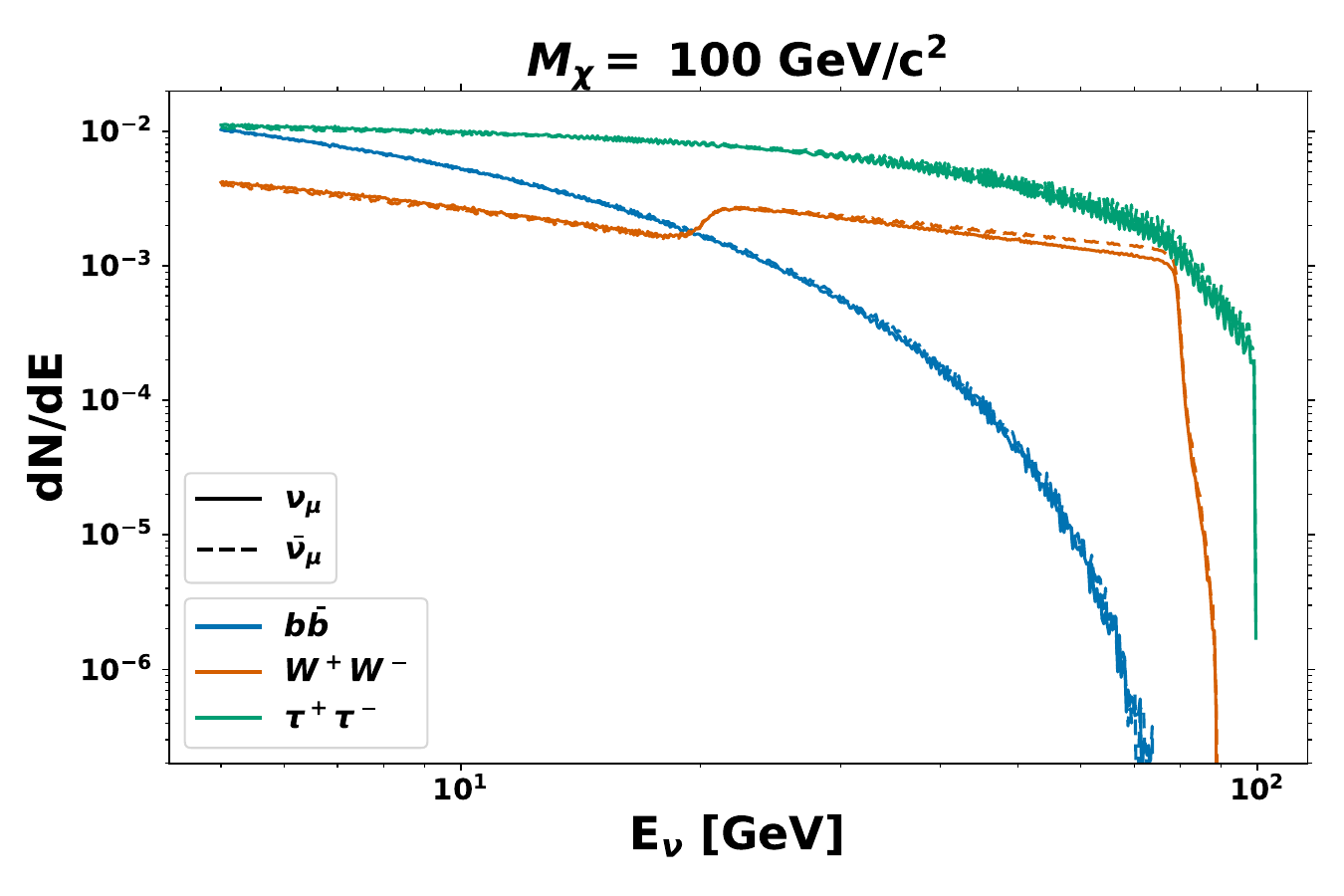}
    \caption{}
\end{subfigure}
\hfill
\begin{subfigure}{0.46\textwidth}
    \centering
    \includegraphics[width=\linewidth]{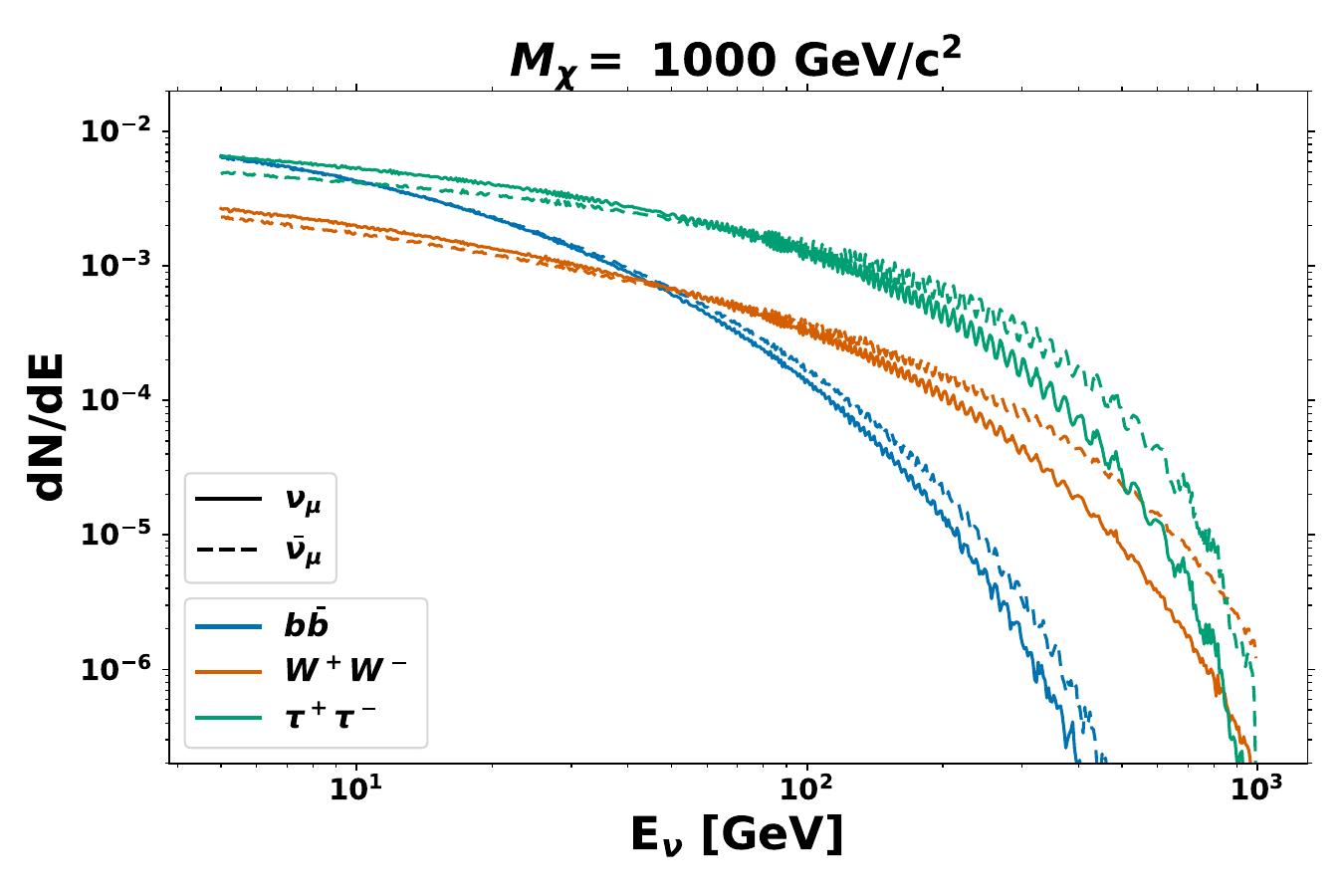}
    \caption{}
\end{subfigure}

\vspace{0.5cm}

\begin{subfigure}{0.46\textwidth}
    \centering
    \includegraphics[width=\linewidth]{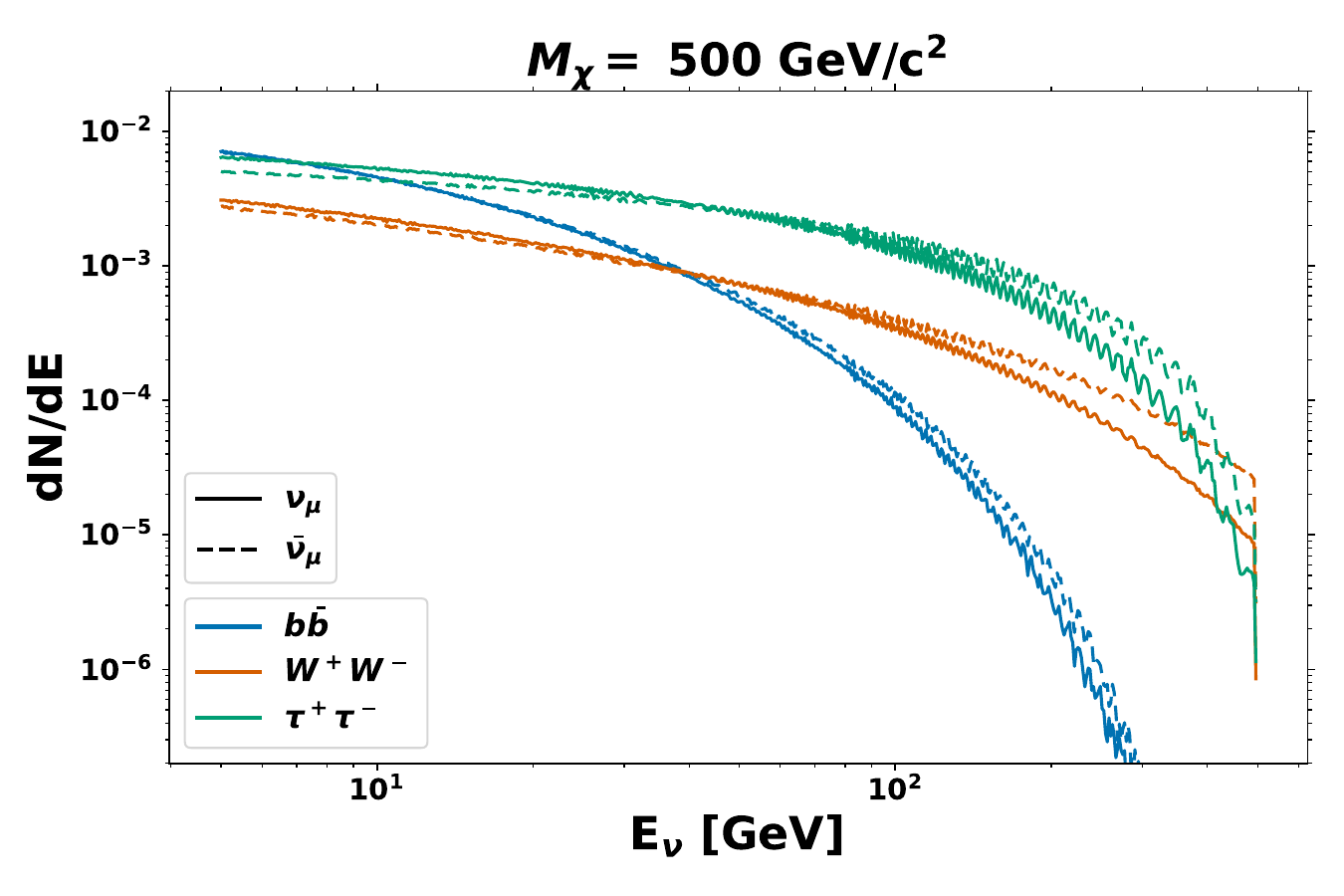}
    \caption{}
\end{subfigure}
\hfill
\begin{subfigure}{0.46\textwidth}
    \centering
    \includegraphics[width=\linewidth]{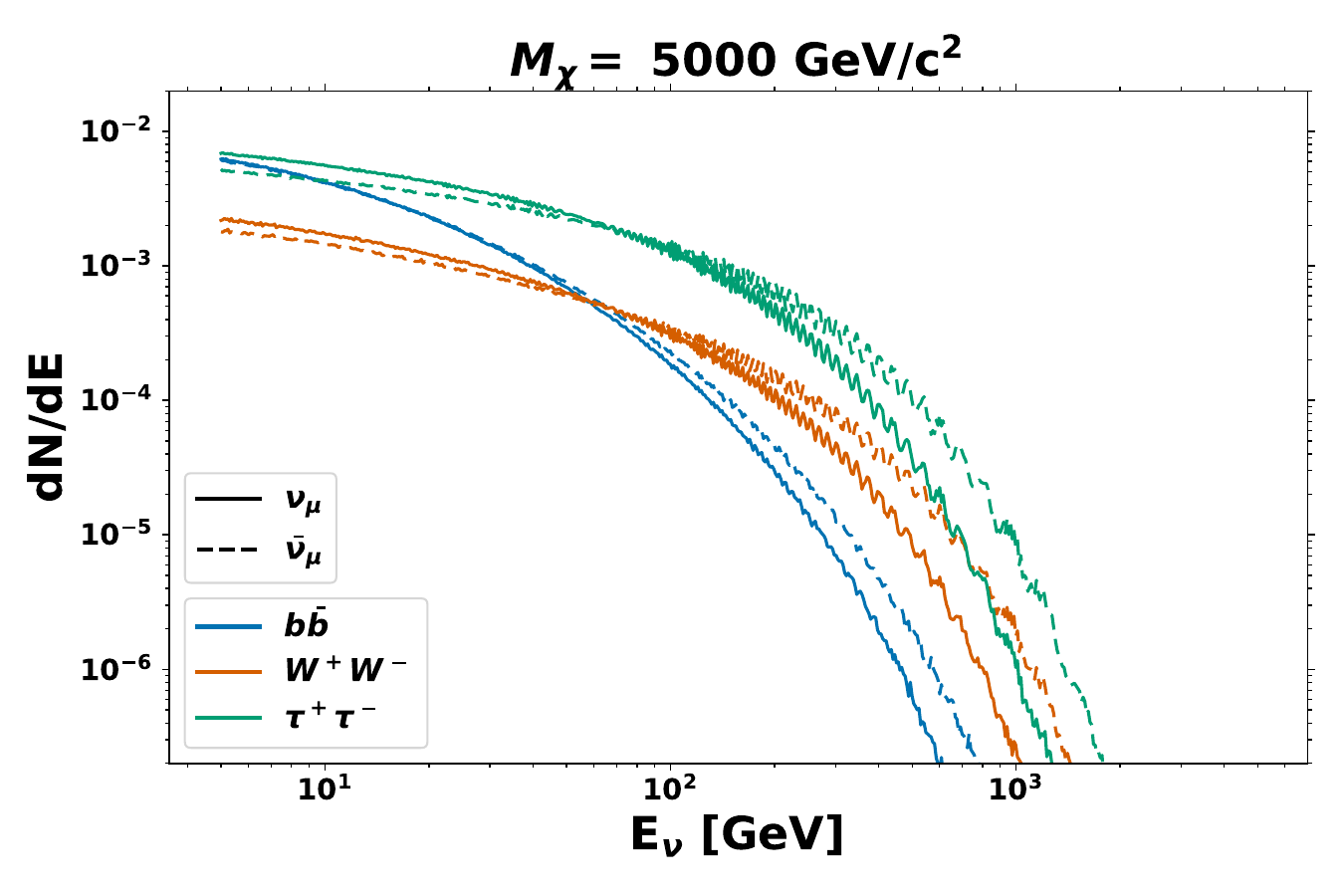}
    \caption{}
\end{subfigure}

\caption{Muon neutrino (solid lines) and muon antineutrino (dashed lines) spectra at the Earth surface for the three channels considered ($W^+W^-$ in red, $\tau^+\tau^-$ in green and $b\bar b$ in blue), and for four WIMP masses: (a) 100 GeV/c$^2$, (b) 1 TeV/c$^2$, (c) 500 GeV/c$^2$, (d) 5 TeV/c$^2$. Analogous spectra have been produced for all the DM masses considered in this work and for all neutrino flavours and decay channels. The wiggles in the high-energy tails are due to neutrino oscillations. Neutrinos from the $\tau^+\tau^-$ channel are also affected by $\tau$ regeneration in the Sun.}
\label{fig:spectra}

\end{figure}

The Sun is located at the periphery of the Milky Way and moves through the Galactic DM halo. This halo is assumed to be populated by single WIMP particle species, with an average local density of 0.3 GeV/cm$^3$. The motion of the Sun in the halo may cause WIMPs to pass through the Sun and scatter off its nuclei, losing enough momentum to become gravitationally trapped \cite{wimpsun}. 
WIMPs, once trapped in the core of the Sun, may annihilate into SM particles, which can produce neutrinos by decaying into or interacting with other particles \cite{ndecay}. 
These SM particles could be leptons, such as $\tau$, gauge bosons, such as $W$, or quarks, such as $b$. In the first case, due to the conservation of the lepton flavour number, a neutrino is always emitted \cite{sun1}. 
The \textit{W} boson can decay leptonically, producing neutrinos, with a $32.6\%$ branching ratio \cite{pdg}, and its hadronic decays can produce particle cascades which in turn yield neutrinos. 
Finally, heavy quarks can produce neutrinos through charged-current (CC) weak interactions, or can generate hadronic showers from which neutrinos can emerge \cite{pdg}.
These neutrinos will then travel through the Sun. The fraction of neutrinos absorbed in the Sun is negligible below a TeV and increases with energy, distorting the neutrino spectrum accordingly. Once the neutrinos exit the Sun, they traverse space affected only by neutrino oscillations. Those neutrinos intersecting the solid angle covered by the Earth may then be detected. \\
Assuming that WIMPs are stable, their density in the core of the Sun will increase with time because of gravitational captures; on the other hand, WIMP-WIMP annihilation should decrease their number, proportionally to the square of the DM density. The age of the Sun is long enough to allow for an equilibrium between the DM capture and annihilation rates to be reached \cite{sunmodel}. 
Under this assumption of equilibrium, the measurement of the annihilation rate, via the detection of the annihilation products, also yields an estimate of the capture rate. Since the composition and properties of the Sun are well-described by the Solar Standard Model \cite{sunmodel}, this rate estimate can be used to infer the DM scattering cross section. \\
The WimpSim software \cite{wimpsim} has been used to simulate the self-annihilation of dark matter particles in the Sun, and to obtain the energy spectrum of resulting neutrinos at Earth; this simulation also accounts for the effects of neutrino oscillations along the neutrino path from the Sun to the Earth, only affected by neutrino oscillations. The WIMP-WIMP annihilation channels considered in this study are $b \bar b$, $\tau^+ \tau^-$ and $W^+ W^-$, assuming that the DM annihilation branching ratio is $100\%$, in each case. By choosing these three channels, it is possible to study DM annihilation in the Sun spanning from the hardest neutrino spectra ($W^+ W^-$ and $\tau^+ \tau^-$) to the softest one ($b \bar b$).
Examples of the neutrino and antineutrino spectra at the Earth surface for dark matter masses of 100, 500, 1000 and 5000\,GeV/c$^2$ are shown in Figure~\ref{fig:spectra}. The dark matter mass values considered in this work are, in units of GeV/c$^2$: 35, 50, 80, 100, 150, 200, 250, 350, 500, 750, 1000, 1500, 2000, 3000, 5000, 7000, 10000.
Neutrino spectra have been computed for all neutrino and antineutrino flavours for each decay channel and for all the above DM masses.

\section{The ANTARES detector and datasets}
\label{sec:antares}

The ANTARES neutrino telescope \cite{antares} was a large-volume neutrino detector located in the Mediterranean Sea, 40\,km from the French town of Toulon, anchored at a depth of $\sim$ 2500\,m. It was the first undersea neutrino telescope, starting its operations in full configuration in 2007 and ending data taking in February 2022. During these years, the detector has produced a large variety of searches in the fields of neutrino physics and astrophysics, multimessenger astronomy, and searches for physics beyond the SM \cite{antaresLast}.\\
The neutrino detection strategy in ANTARES was based on the collection of the Cherenkov light induced in water by the secondary charged particles produced in the interaction of high-energy neutrinos with matter. ANTARES was sensitive to neutrinos of all flavours with energies above a few tens of GeV.
The detector structure consisted of a three-dimensional array of \textit{optical modules} distributed along 12 vertical detection lines of 450\,m in length. Each line had 25 \textit{floors}, with a vertical spacing of 14.5\,m. On each floor, three optical modules were installed, spaced $120^{\circ}$ apart and looking 45$^{\circ}$ downwards. Each optical module consisted of a pressure-resistant glass sphere hosting a $10^{\prime\prime}$ photomultiplier tube. 
The lines were held vertical by buoys located at the top.
The separation between lines ranged from 60 to 75 m. Each line was deployed using a ship and connected to a \textit{junction box} using a remotely operated submarine vehicle. The junction box was connected to shore via an electro-optical cable for power supply and data transfer.
The detector also included several calibration systems \cite{instrumentation}. \\
Data have been collected by ANTARES almost continuously from February 2007 to February 2022. All data sent to shore were processed by online filtering algorithms and interesting candidate events were stored for subsequent analyses. A preliminary data-quality selection is applied on all data stored for offline analysis, which allows excluding data collected for calibration purposes or during periods with unstable data acquisition conditions. After this preselection, a total of 12.47 years (4551 days) of effective detector livetime becomes available for the analysis.

\subsection{Event reconstruction strategies}

All triggered events collected during the ANTARES livetime have been processed by different algorithms which reconstruct the arrival direction and energy of the charged leptons inducing the detected Cherenkov photons.
To optimise the detector performance in determining the properties of the incoming neutrinos at different energy ranges, three different reconstruction strategies have been used.
Each reconstruction strategy is associated with a quality parameter, which quantifies the compatibility of the reconstructed event with the hypothesis of a muon track.
To remove atmospheric muons, which constitute the main source of background, up-going events are selected, as only neutrinos can cross the Earth without being significantly absorbed. After this selection, the remaining background is mostly due to atmospheric neutrinos originated from the opposite side of the Earth. This quasi-isotropic background can be significantly reduced by selecting reconstructed events originating from a narrow sky region around the expected cosmic neutrino source, here the Sun. \\
A selection based on the quality parameters provided by the reconstruction algorithms is performed to further increase the signal over background ratio against the presence of wrongly reconstructed atmospheric muons. 
For events that are properly reconstructed by more than one of the three strategies, only the output of the algorithm providing the best sensitivity in the search for neutrinos from DM annihilation in the Sun is selected, independently for each of the analysed WIMP masses and DM annihilation channel. In the following there is a description of the three methods.

\paragraph{\textbf{\textit{Method 1}}} It applies deep learning models to reconstruct events with hits detected only in one line, enabling the estimation of the event direction, geometry, and energy along with the corresponding uncertainties. This reconstruction strategy is targeted, in particular, towards low-energy neutrinos, which are more likely to hit only one line, boosting the sensitivity to low dark matter masses and to neutrinos with energy below $\sim$100 GeV \cite{nnfit}.
The uncertainty on the direction, $\sigma_{\Omega}$, will be used as quality parameter for the selection cut using this method. Cut values ($\sigma_{\Omega} < \sigma_{\Omega}^{\textrm{cut}}$) are investigated using MC simulations, for $\sigma_{\Omega}^{\textrm{cut}}$ ranging from $20^{\circ}$ to $28^{\circ}$ with a step of $1^{\circ}$.

\paragraph{\textbf{\textit{Method 2}}} It is a $\chi^2$-like fit and is most effective for the reconstruction of neutrinos at intermediate energies \cite{bbfit}. 
The $\chi^2$-like fit is an online reconstruction algorithm developed for real-time analyses, involving either sending neutrino alerts or processing external triggers \cite{alerts}. It employs a quality function based on the residuals between expected and measured photon arrival times, and incorporates a correction for light absorption.
For this method, the $\chi^2$ value is used as quality parameter for the selection cut. A scan over this parameter is performed for atmospheric muon contamination rates ranging from $90\%$ ($\chi^2 < 1.7$) to $30\%$ ($\chi^2 < 0.6$) in steps of 0.1.\\

\paragraph{\textbf{\textit{Method 3}}} It is a maximum likelihood method, which is optimal for the reconstruction of neutrinos at energies larger than $\sim$500\,GeV \cite{aafit}. 
This approach provides excellent angular resolution and improved detection efficiency at high energies. It exploits the arrival time and charge information of the hits, reconstructing the muon track through successive fitting procedures of increasing complexity and progressively refined hit selections. The final stage is a maximum-likelihood fit, where the likelihood function $\Lambda$ is defined in terms of the probability density of the time residuals.
Another parameter combined with $\Lambda$ is $\beta$ which represents the angular error estimator. Both quantities are used as quality parameters for the selection cuts.
A 2D scan is performed for $\Lambda > \Lambda^{\textrm{cut}}$ with $\Lambda^{\textrm{cut}}$ from $-5.7$ to $-5.3$ in steps of 0.1 and for $\beta$ cut values ($\beta < \beta^{\textrm{cut}}$) are investigated for $\beta^{\textrm{cut}}$ ranging from 0.7$^{\circ}$ to 1$^{\circ}$ with steps of 0.1.\\

\section{The analysis strategy}
\label{sec:analysis}

The goal of the search for a DM signal from the direction of the Sun with the ANTARES neutrino telescope is the observation of a significant excess of events above the astrophysical and atmospheric backgrounds. Given the optimal directional reconstruction achieved for events in which a muon passes through the detector, this analysis considers events reconstructed as up-going tracks. \\
In the ANTARES Collaboration, the expected rate of detected events is modelled using a run-by-run \cite{ANTARESMC} simulation approach, reproducing the time-dependent evolution of the detector, to assess the impact of the different selection cuts. For the case of this analysis, the background is estimated by scrambling the time information of the events in the dataset, taking advantage of the fact that the Sun is a \textit{moving} source. \\
The analysis optimisation is blinded, as the real arrival direction of neutrinos in data is only used after all event selection cuts have been fixed on simulations and scrambled data.
The search strategy is based on the following procedure: first, a signal hypothesis is defined (neutrinos produced by solar WIMP annihilation); then, atmospheric muon and neutrino backgrounds are characterised. 
Astrophysical backgrounds have been previously constrained \cite{daniel} to such a level that allows us to neglect them.
Event-level observables allowing discriminating signal from background, such as angular separation from the Sun or reconstruction quality parameters, are used to select signal event candidates.  
The analysis relies on an extended maximum-likelihood approach, in which each event contributes as a weighted combination of signal and background probability density functions (PDFs), while the total number of signal events is treated as a free parameter. The maximum-likelihood function used is the following: 
\begin{equation}
    \ln \mathcal{L}(n_s) = \sum^N_{i=1} \ln [n_s S(\Psi_{\odot, \textrm{i}}, \beta_\textrm{i}, N_{\textrm{hits, i}}) + n_b B(\Psi_{\odot, \textrm{i}}, \beta_\textrm{i}, N_{\textrm{hits, i}})] - (n_s + n_b)
\end{equation} 
where $S(\Psi_{\odot, i}, \beta_\textrm{i}, N_{\textrm{hits, i}})$ and $B(\Psi_{\odot, \textrm{i}}, \beta_\textrm{i}, N_{\textrm{hits, i}})$ represent the signal and the background PDFs for an event at an angular distance $\Psi_{\odot, i}$ from the Sun, while $n_s$ and $n_b$ are the numbers of signal and background events, respectively. $N_{\textrm{hits, i}}$ and $\beta_\textrm{i}$ are the number of hits in the event and its angular error estimate, respectively, and contribute to the discrimination between signal and background given the different shape of their distributions under these two hypotheses. \\
Sensitivities in terms of number of signal events are set in a procedure of hypothesis testing. The background-only hypothesis $H_0$ is compared against an alternative, signal+background hypothesis $H_s$.
The chosen Test Statistic (TS) is the likelihood-ratio test, as first proposed by Neyman and Pearson \cite{NP}. In this approach, the TS is the ratio between the likelihood computed for the set of parameters that maximise it, and the likelihood computed for the parameters as constrained by the null hypothesis.
The TS distributions for the null (background-only) hypothesis and for different numbers of signal events are evaluated by generating pseudo-experiments (PEXs), sampling the PDFs with a given value of injected signal events. \\
Since in this analysis the parameter of interest is the number of signal events ($n_s$), the TS can be expressed as:
\begin{equation}
    TS = \log_{10} \left( \frac{\mathcal{L}(n_s = \hat{n}_s)}{\mathcal{L}(n_s = 0)} \right) 
    \label{eq:ts}
\end{equation}\\
with $\hat{n}_s$ indicating the best fit value.
For every WIMP mass, annihilation channel, reconstruction strategy,
set of cuts and number of injected signal events, at least ten thousands PEXs are generated to ensure a smooth determination of the TS distributions.\\
The optimum final selection cuts are determined using the TS distributions computed with the pseudoexperiments. The best sensitivity to the flux of neutrinos from WIMP annihilation ($\bar {\Phi}_{\nu + \bar{\nu}} ^{90\%}$) is defined as:
\begin{equation}
    \bar {\Phi}_{\nu + \bar{\nu}} ^{90\%} = \frac{\bar {\mu}_{90\%}}{\bar {A}_{\textrm{cc}}(M_{\textrm{WIMP}}) \cdot T_{\textrm{eff}}} ,
    \label{eq:sensy}
\end{equation}
This expression is obtained from the TS distributions extracting the average upper limit on the number of signal events at $90\%$ confidence level ($\bar {\mu}_{90\%}$), divided by the total livetime of the detector (in this case $T_{\textrm{eff}}= 4551$\,days), and acceptance $\bar {A}_{\textrm{cc}}(M_{\textrm{WIMP}})$ for a given WIMP mass and annihilation channel.
The detector acceptance for the interactions of neutrinos originated by DM annihilation in the Sun can be expressed as 
\begin{equation}
    \bar {A}_{\textrm{cc}}(M_{\textrm{WIMP}}) =  \frac{\sum_{j=\nu,\bar {\nu}}
    \left( \int_{5\,\textrm{GeV}}^{M_{\textrm{WIMP}}} A_{\textrm{eff}}^j (E_j) 
    \frac{dN_j}{dE_j} dE_j\right)}
    {\int_{5\,\textrm{GeV}}^{M_{\textrm{WIMP}}} 
    \left( \frac{dN_{\nu}} {dE_{\nu}} dE_{\nu} + 
    \frac{dN_{\bar {\nu}}} {dE_{\bar {\nu}}} dE_{\bar {\nu}} \right) } ,
    \label{eq:acc1}
\end{equation}
\noindent where $dN_{\nu, \bar {\nu}}/dE_{\nu, \bar {\nu}}$ is the energy spectrum of the (anti-)neutrinos at the surface of the Earth, simulated using WimpSim \cite{wimpsim} for each of the considered channels, $b \bar b$, $\tau^+ \tau^-$ and $W^+ W^-$. The integrals start from the value of $5$\,GeV, which is a conservative value below the energy threshold of the telescope.
$A_{\textrm{eff}}^{\nu, \Bar{\nu}} (E_{\nu, \Bar{\nu}})$ is the effective area of the ANTARES detector as a function of the (anti-)neutrino energy, depending on the selection cuts.\\
The event reconstruction method and the corresponding selection cuts resulting from the optimisation are reported in Table~\ref{tab:cuts} for each DM mass range, and DM annihilation channel. 

\begin{table}[!h]
\caption{Chosen selection cuts and event reconstruction strategy that optimise the ANTARES sensitivity to neutrinos from WIMP annihilations, for each channel and WIMP mass range.}
\begin{center}
\begin{tabular}{ccccc}
Channel & WIMP mass & Strategy & Best cut on \\
 & (GeV/c$^2$) & & quality parameter & \\
\hline
\hline
$b\bar{b}$ & 35 $\le$ M $\le$ 100  & method 1 & $\sigma_{\Omega}$ $<$ 24$^{\circ}$   \\
$b\bar{b}$ & 150 $\le$ M $\le$ 350  & method 2 & $\chi^2$ $\le$ 1.4   \\
$b\bar{b}$ & M $\ge$ 500  & method 3 & $\Lambda$ $\ge$ $-5.5$, $\beta \le 0.8^{\circ}$ \\
\hline
$\tau^+\tau^-$ & 35 $\le$ M $\le$ 50  & method 1 & $\sigma_{\Omega}$ $<$ 24$^{\circ}$   \\
$\tau^+\tau^-$ & 80 $<$ M $\le$ 100  & method 2  & $\chi^2$ $\le$ 1.4  \\
$\tau^+\tau^-$ & M $\ge$ 150  & method 3 & $\Lambda$ $\ge$ $-5.5$, $\beta \le 0.8^{\circ}$\\
\hline
$W^+W^-$ &  M $= 100$  & method 2  & $\chi^2$ $\le$ 1.4  \\
$W^+W^-$ & M $\ge$ 150  & method 3 & $\Lambda$ $\ge$ $-5.5$, $\beta \le 0.8^{\circ}$  \\
\hline
\hline
\end{tabular}
\end{center}
\label{tab:cuts}
\end{table}

\section{Results}
\label{sec:results}

\begin{figure}[!h]
\begin{center}
\includegraphics[width=1.05\textwidth]{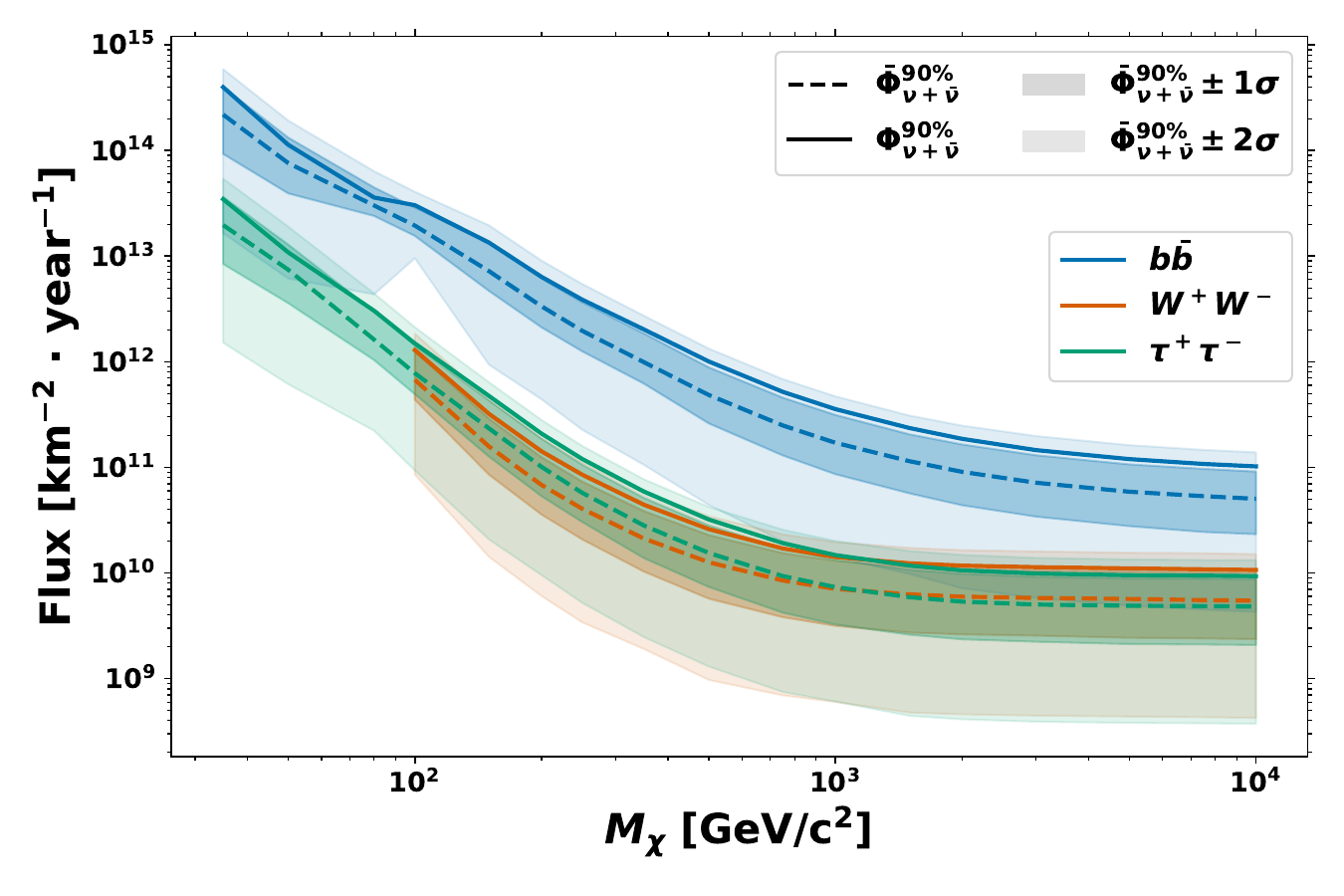}
\caption{Upper limits at 90$\%$ CL ($\Phi_{\nu + \bar{\nu}}^{90\%}$, solid lines) on the flux of neutrinos from DM annihilation in the Sun, using 4551 livetime days of ANTARES data (2007-2022) as a function of the dark matter mass. The three different channels are reported in different colors: $b \bar{b}$ in blue, $\tau^+ \tau^-$ in green and $W^+ W^-$ in red. The sensitivity fluxes ($\bar{\Phi}_{\nu + \bar{\nu}}^{90\%}$) are reported with dashed lines, showing as well the $\pm 1 \sigma$ and $\pm 2 \sigma$ bands.}
\label{fig:upperlim}
\end{center}
\end{figure}

\begin{figure}[!h]
\begin{center}
\includegraphics[width=1\textwidth]{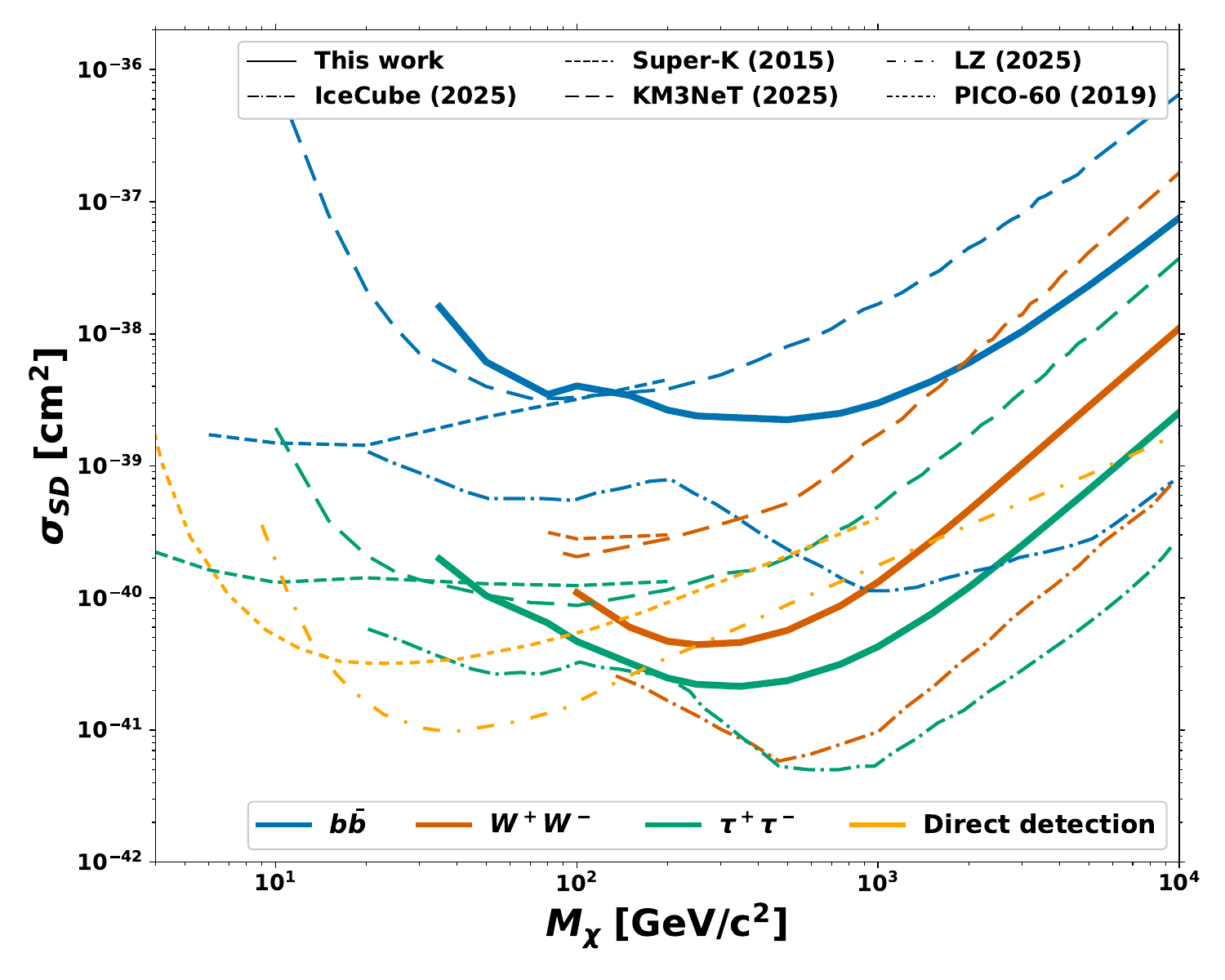}
\caption{Upper limits at $90\%$ CL on the spin–dependent WIMP–nucleon scattering cross section as a function of WIMP mass for the $b \bar{b}$ (blue), $\tau^+ \tau^-$ (green) and $W^+ W^-$ (red) channels. Corresponding 90$\%$ CL upper limits given by other experiments are also included: IceCube \cite{ICSun}, SuperKamiokande \cite{SKSun}, KM3NeT \cite{km3sun}, PICO-60 \cite{PICOSun}, Lux-Zeplin \cite{LZ}.}
\label{fig:SDall}
\end{center}
\end{figure}

\begin{figure}[!h]
\begin{center}
\includegraphics[width=1\textwidth]{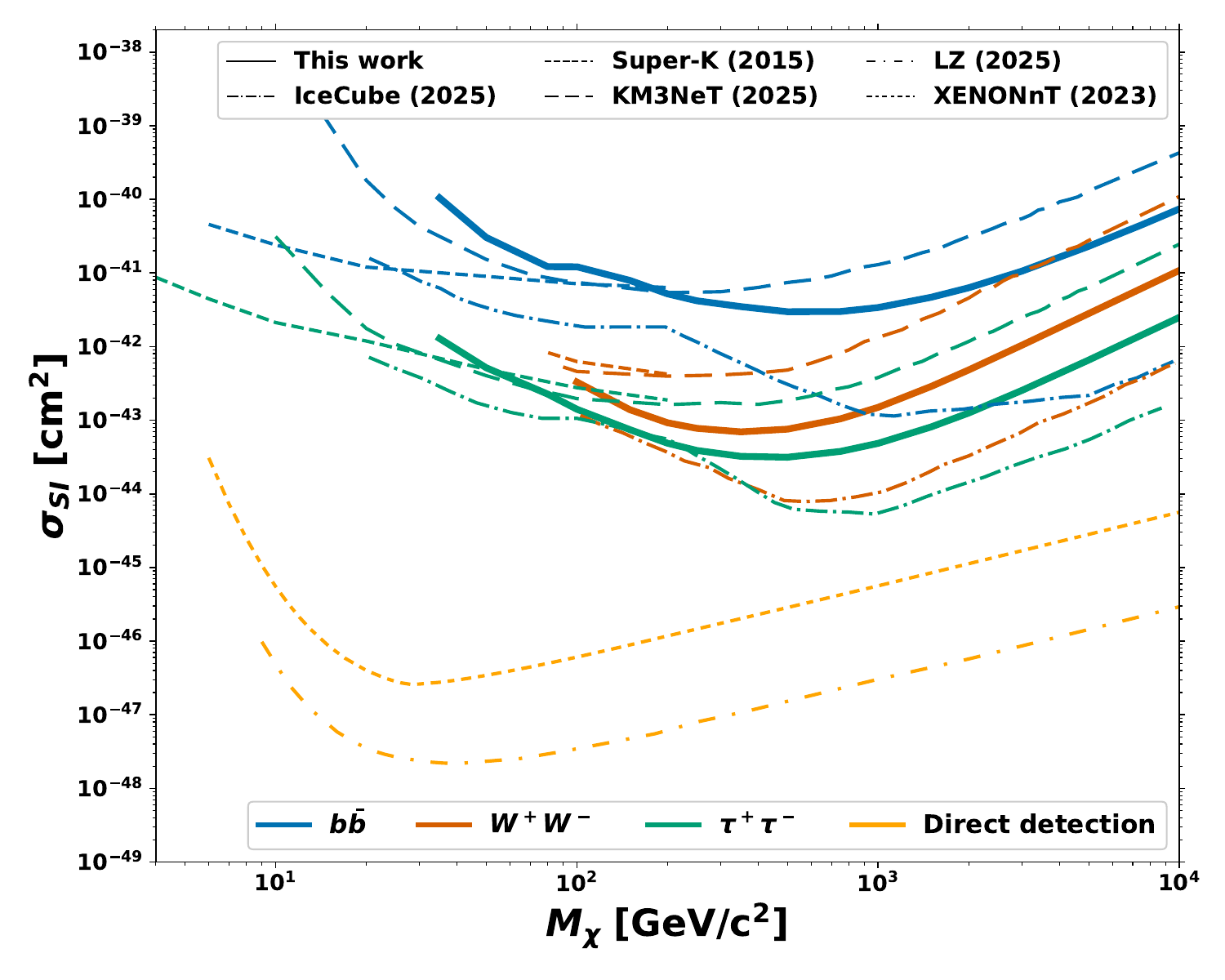}
\caption{Upper limits at $90\%$ CL on the spin–independent WIMP–nucleon scattering cross section as a function of WIMP mass for the $b \bar{b}$ (blue), $\tau^+ \tau^-$ (green) and $W^+ W^-$ (red) channels. Corresponding 90$\%$ CL upper limits given by other experiments are also included: IceCube \cite{ICSun}, SuperKamiokande \cite{SKSun}, KM3NeT \cite{km3sun}, XENONnT \cite{XenonSun}, Lux-Zeplin \cite{LZ}.}
\label{fig:SIall}
\end{center}
\end{figure}

Once the optimal selection cuts are chosen, the unblinding procedure consists of applying the analysis described in Section 4, with the corresponding optimised cuts, to unscrambled data. The observed values of the test statistic TS$_{\textrm{obs}}$ are then compared to the background-only TS distribution to extract a best fit on the number of observed events from the signal, as reported in Table~\ref{tab:numbers}. The statistical significance of the observation is evaluated using a one-sided p-value, defined as the probability to measure a TS greater than or equal to TS$_{\textrm{obs}}$ under the background-only hypothesis.
For all tested WIMP masses and annihilation channels, the contribution from the DM signal is compatible with the background expectations within statistical uncertainties.
Thus, limits on the neutrino flux from the Sun  produced by dark matter annihilation were derived for the three DM annihilation channels, and are shown in Figure~\ref{fig:upperlim}. The size of the bands which accompany the sensitivities accounts for the estimated systematic uncertainties, obtained considering both the astrophysical assumptions made in the estimation of the DM signal and those made in the estimation of the detector performance \cite{unc}, for the signal as well as for the background. 
Uncertainties in the signal estimation come from uncertainties on the neutrino-nucleon cross section, on the neutrino oscillation parameters, and on the WimpSim assumptions for the computation of the neutrino spectra from DM annihilation. On the other hand, the detector-related uncertainties are connected with the uncertainties on the estimation of the background from optimisation and analysis procedure, and on the detector efficiency as obtained from Monte Carlo simulations.
The combination of the two contributions gives a global uncertainty of about 20$\%$ \cite{unc}.\\
The resulting flux limits can be converted into WIMP-nucleon interaction cross section upper limits on the basis of the input parameters used by WimpSim in the simulation of the DM annihilation signal provided by WimpSim \cite{wimpsim, crossconv}. 
Several parameters enter this estimation: the equilibrium between WIMPs annihilation and capture processes which has been reached in the Sun, the number of neutrinos produced for each annihilation process, the mean square distance between Sun and Earth, and the effects of neutrino oscillations.
The upper limits on the WIMP-proton scattering cross section obtained in this analysis, assuming spin dependent interactions, are shown in Figure~\ref{fig:SDall}, together with the results from the IceCube \cite{ICSun} and Super-Kamiokande \cite{SKSun} neutrino detectors, and from PICO-60 \cite{PICOSun} and Lux-Zeplin \cite{LZ}, the direct detection experiment that have given, so far, the most stringent upper limits. Results from the next-generation neutrino observatory in the  Mediterranean Sea, KM3NeT \cite{km3}, currently under construction and already taking data, are also included \cite{km3sun}. Figure~\ref{fig:SIall} shows the upper limits for the spin independent case, also compared with results from other neutrino detectors, and from the direct detection experiment XENONnT \cite{XenonSun}.
These results improve upon the ones achieved by the last ANTARES publication \cite{SunC} by factors between two and three thanks to the increased livetime and improved reconstruction of low energy events.
In the DM mass region above 100\,GeV/c$^2$, the indirect detection experiments produce better results than the direct ones for the spin-dependent case, when assuming that DM annihilates mainly through the channels considered here. For lower DM masses, Super-Kamiokande performed better than ANTARES. 
Differences between ANTARES and IceCube upper limits can be partially explained by the tools used to produce the neutrino spectra: for ANTARES, WIMPSim has been used, while IceCube relied on a new tool, called $\chi aron$ \cite{caron}, which includes a complete calculation of the electroweak effect that can result in harder neutrino spectra, improving up to an order of magnitude the resulting limits. \\
The results obtained by neutrino telescopes complement each other, due to the different detector locations, fields of view, energy ranges, and analysis strategies, thus leading to robust upper bounds on the DM annihilation cross section in the Sun over a large DM mass range. For the spin-independent case, direct detection experiments produce more stringent limits but, since the systematic uncertainties are different, the results obtained with indirect searches can provide useful information to study DM models.

\section{Summary and conclusions}
\label{sec:conclusions}

The analysis presented here uses 15 years (2007-2022) of data collected by the ANTARES neutrino telescope to constrain the DM-nucleon scattering cross section in the Sun for DM masses ranging from 35\,GeV/c$^2$ to 10\,TeV/c$^2$.  
No significant excess of events from the direction of the Sun has been observed and limits on the cross section have been set for both the spin-dependent and the spin-independent  WIMP-nucleon cross sections. 
The results obtained in this work improve by a factor between two and three upon those previously published by ANTARES, which used a dataset covering the  2007-2012 period \cite{SunC}. These upper limits are competitive with those obtained in both direct and indirect searches for DM masses greater than 100\,GeV/c$^2$. In the case of spin-dependent WIMP-nucleon scattering, the limits obtained in the above-mentioned DM mass range are more stringent than those from direct detection experiments, similarly to what has been obtained in other indirect searches.

\begin{table}{}
\caption{Summary of the unblinding results in the search for neutrinos from WIMP annihilation in the Sun. The five columns show the values of the number of total events (N$_{\textrm{tot}}$), number of fitted signal events ($\hat{n}_{s}$), the corresponding p-value, the acceptance and the $\mu_{90}$, for all studied masses and channels.}
\begin{center}
\scriptsize
\begin{tabular}{ccccccc}
WIMP mass & Channel & N$_{\textrm{tot}}$ & $\hat{n}_{s}$ & p-value & Acceptance & $\mu_{90}$  \\
(GeV/c$^2$) &  & & & & (m$^2$) & \\
\hline
\hline
35 & $b\bar{b}$ & 361  & 0.6  & 0.14 & 1.9$\times 10^{-09}$ & 9.6  \\
 & $\tau^+\tau^-$ & 361 & 0.5  &  0.17 & 1.8$\times 10^{-08}$  & 8.0   \\
\hline
50 & $b\bar{b}$ & 1179  & 0.0 & 0.23  & 9.8$\times 10^{-09}$  & 13.8  \\
 & $\tau^+\tau^-$ & 1179 & 0.0  &  0.24 & 9.2$\times 10^{-08}$  & 12.4   \\
\hline
80 & $b\bar{b}$ & 3055 & 0.0 & 0.32 & 4.4$\times 10^{-08}$  & 19.8  \\
 & $\tau^+\tau^-$ & 7976 & 5.9  &  0.14  & 3.9$\times 10^{-07}$  & 14.3  \\
\hline
100 & $b\bar{b}$ & 3987 & 0.0  & 0.09  & 7.8$\times 10^{-08}$  & 29.4  \\
 & $\tau^+\tau^-$ & 7976 & 6.0  &  0.12  & 7.7$\times 10^{-07}$  & 14.3  \\
 & $W^+W^-$ & 7976 & 5.9  &  0.13 & 8.9$\times 10^{-07}$  & 14.3   \\
\hline
150 & $b\bar{b}$ & 7976 & 6.0  &  0.14 & 8.6$\times 10^{-08}$  & 14.4\\
 & $\tau^+\tau^-$ &  22056 & 5.8 &  0.10 & 2.3$\times 10^{-06}$  & 13.7   \\
 & $W^+W^-$ & 22056 & 5.6  &  0.10 & 3.4$\times 10^{-06}$  & 13.5   \\
\hline
200 & $b\bar{b}$ & 7976 & 6.1  &  0.12 & 1.8$\times 10^{-07}$  & 14.5   \\
& $\tau^+\tau^-$ &  22056 & 5.3 &  0.09 & 5.0$\times 10^{-06}$  & 13.1    \\
& $W^+W^-$ & 22056 & 5.2  &  0.09  & 7.3$\times 10^{-06}$  & 12.9  \\
\hline
250 & $b\bar{b}$ & 7976 & 6.2  &  0.12 & 3.0$\times 10^{-07}$  & 14.5    \\
 & $\tau^+\tau^-$ &  22056 & 5.1 &  0.09 & 8.4$\times 10^{-06}$  & 12.6   \\
 & $W^+W^-$ & 22056 & 4.9  & 0.08  & 1.2$\times 10^{-06}$  & 12.5  \\
\hline
350 & $b\bar{b}$ & 7976 & 6.3  &  0.10  & 5.7$\times 10^{-07}$  & 14.5    \\
 & $\tau^+\tau^-$ &  22056 & 4.6 &  0.08  & 1.6$\times 10^{-06}$  & 11.9  \\
 & $W^+W^-$ & 22056 & 4.3  &  0.08  & 2.1$\times 10^{-06}$  & 11.6   \\
\hline
500 & $b\bar{b}$ & 22056 & 5.7  &  0.09  & 1.1$\times 10^{-06}$  & 13.7   \\
 & $\tau^+\tau^-$ & 22056 & 4.0  &  0.09  & 2.8$\times 10^{-05}$  & 11.1  \\
 & $W^+W^-$ & 22056 & 3.8  &  0.09  & 3.3$\times 10^{-06}$  &  10.7  \\
\hline
750 & $b\bar{b}$ & 22056 & 5.3  &  0.09 & 2.0$\times 10^{-06}$  & 13   \\
 & $\tau^+\tau^-$ & 22056 & 3.5  & 0.09 & 4.3$\times 10^{-05}$  & 10.2   \\
 & $W^+W^-$ & 22056 & 3.2  &  0.10  & 4.6$\times 10^{-06}$  &  9.8\\
\hline
1000 & $b\bar{b}$ & 22056 & 5.0  &  0.09  & 2.8$\times 10^{-06}$  & 12.6   \\
 & $\tau^+\tau^-$ & 22056 & 3.5  &  0.10  & 5.3$\times 10^{-05}$  & 9.7  \\
 & $W^+W^-$ & 22056 & 3.0  &  0.10 & 5.3$\times 10^{-06}$  &  9.4 \\
\hline
1500 & $b\bar{b}$ & 22056 & 4.6  &  0.09 & 4.1$\times 10^{-06}$  & 12.0   \\
 & $\tau^+\tau^-$ &  22056 & 2.9 &  0.10 & 6.3$\times 10^{-06}$  & 9.2   \\
 & $W^+W^-$ & 22056 & 2.7  &  0.11 &  5.9$\times 10^{-06}$ & 9.0   \\
\hline
2000 & $b\bar{b}$ & 22056 & 4.4  &  0.09 & 5.0$\times 10^{-06}$  &  11.7    \\
 & $\tau^+\tau^-$ & 22056 & 2.8  &  0.11 & 6.8$\times 10^{-05}$  & 9  \\
 & $W^+W^-$ & 22056 & 2.7  &  0.11 & 6.1$\times 10^{-06}$  & 8.9 \\
\hline
3000 & $b\bar{b}$ & 22056 & 4.1  &  0.09  & 6.2$\times 10^{-06}$  &  11.3  \\
 & $\tau^+\tau^-$ & 22056 & 2.7  &  0.11  & 7.2$\times 10^{-05}$  & 8.8  \\
 & $W^+W^-$ & 22056 & 2.7  &  0.11 & 6.3$\times 10^{-06}$  &  8.9   \\
\hline
5000 & $b\bar{b}$ & 22056 & 3.9  &  0.09 & 7.4$\times 10^{-06}$  &  11   \\
 & $\tau^+\tau^-$ & 22056 & 2.6  &  0.11 & 7.4$\times 10^{-05}$  &  8.7 \\
 & $W^+W^-$ &  22056 & 2.6 &  0.11 & 6.4$\times 10^{-06}$  & 8.8 \\
\hline
7500 & $b\bar{b}$ & 22056 & 3.8  &  0.09   & 8.0$\times 10^{-06}$  & 10.8  \\
 & $\tau^+\tau^-$ & 22056 & 2.6  &  0.11  & 7.4$\times 10^{-05}$  & 8.7  \\
 & $W^+W^-$ &  22056 & 2.6  &  0.11  &  6.5$\times 10^{-06}$ &  8.8 \\
\hline
10000 & $b\bar{b}$ &  22056 & 3.7 &  0.09   & 8.4$\times 10^{-06}$  &  10.7 \\
 & $\tau^+\tau^-$ & 22056 & 2.6  &  0.11 & 7.5$\times 10^{-05}$  & 8.7   \\
 & $W^+W^-$ &  22056 & 2.6  & 0.11 & 6.6$\times 10^{-06}$  & 8.8  \\
\hline
\hline
\end{tabular}
\end{center}
\label{tab:numbers}
\end{table}

\section*{Acknowledgements}
The authors acknowledge the financial support of the funding agencies:
Centre National de la Recherche Scientifique (CNRS), Commissariat \`a
l'\'ener\-gie atomique et aux \'energies alternatives (CEA),
Commission Europ\'eenne (FEDER fund and Marie Curie Program),
LabEx UnivEarthS (ANR-10-LABX-0023 and ANR-18-IDEX-0001),
R\'egion Alsace (contrat CPER), R\'egion Provence-Alpes-C\^ote d'Azur,
D\'e\-par\-tement du Var and Ville de La
Seyne-sur-Mer, France;
Bundesministerium f\"ur Bildung und Forschung
(BMBF), Germany; 
Istituto Nazionale di Fisica Nucleare (INFN), Italy;
Nederlandse organisatie voor Wetenschappelijk Onderzoek (NWO), the Netherlands;
Ministry of Education and Scientific Research, Romania;
MICIU for PID2024-156285NB-C41, -C42- C43, funded by MICIU/AEI/10.13039/501100011033 and by FEDER, EU, and for CNS2023-144099; Generalitat Valenciana for CIDEGENT/2020/049, CIDEGENT/2021/23, CIDEIG/2023/20, CIPROM/2023/51 and INNVA1/2024/110 (IVACE+i), and  Fundaci\'{o}n Bancaria La Caixa (ID 100010434), for LCF/BQ/PI25/12100025, Spain;
Ministry of Higher Education, Scientific Research and Innovation, Morocco, and the Arab Fund for Economic and Social Development, Kuwait.
We also acknowledge the technical support of Ifremer, AIM and Foselev Marine
for the sea operation and the CC-IN2P3 for the computing facilities.


\end{document}